\documentstyle [prb,aps,eqsecnum]{revtex}
\newcommand{\be}{\begin{equation}}
\newcommand{\ee}{\end{equation}}
\newcommand{\ba}{\begin{eqnarray}}
\newcommand{\ea}{\end{eqnarray}}
\newcommand{\cdag}{c^{\dag}}
\newcommand{\Palpha}{P_{\alpha}}
\newcommand{\Pdag}{P^{\dag}_{\alpha}}
\newcommand{\Pbeta}{P^{\dag}_{\beta}}
\newcommand{\hqa}{h_{q\alpha}}

\newcommand{\hpb}{h_{p\beta}}
\newcommand{\omegalpha}{\omega_{\alpha}}
\newcommand{\prbee}{Phys. Rev. B }
\newcommand{\prel}{Phys. Rev. Lett.\ }

\begin{document}
\draft
\title{Dynamics of compressible edge and bosonization} 
\author{ J. H. Han and D. J. Thouless}
\address{Department of Physics, Box 351560, University of Washington, 
Seattle, Washington 98195} 

\maketitle


\begin{abstract}
 We work out the dynamics of the compressible edge of the quantum Hall 
system based on the electrostatic model of Chklovskii {\it et al.}.
We introduce a generalized version of Wen's 
hydrodynamic quantization approach to the dynamics of sharp edge and 
rederive Aleiner and Glazman's earlier result of multiple density modes.
Bosonic 
operators of density excitations are used to construct fermions at the 
interface of the compressible and incompressible region. We also analyze 
the dynamics starting with the second-quantized Hamiltonian in the lowest 
Landau level and work out the time development of density operators. 
Contrary to the hydrodynamic results, the density modes are strongly coupled. 
We argue that the coupling suppresses the propagation of all acoustic 
modes, and that the excitations with large wavevectors are subject to decay 
due to coupling to the dissipative acoustic modes.A possible correction to 
the tunneling density of states is discussed.

\end{abstract}

\section{Introduction}
The theory of edge excitations in the fractional quantum hall liquid had 
been first derived by Wen\cite{wen1} based on the gauge invariance of 
the Chern-Simons term for a bounded system. A more intuitive picture of 
the edge excitation was laid out in his subsequent paper\cite{wen2} where 
a classical energy and a classical equation of motion for the 
incompressible liquid is written down, and he stipulated that the 
conjugate variables in the corresponding Lagrangian satisfy the canonical 
commutation relations. In the absence of a more microscopic theory, 
such ``phenomenological quantization" serves as a guideline in our 
understanding of the phenomena and not infrequently yields accurate 
results. Quite a lot of progress in the study of vortex motion in superfluid He 
and superconductors had been made in this way.\cite{haldane,ao} 

Eventually, any phenomenological theory has to be based on the analysis of
the underlying microscopic Hamiltonian.
A good example is the plasma oscillation in a metal where one can show 
that a quantum-mechanical equation of motion of the density operator 
reduces to 
the (semi-)classical one in the regime where RPA is applicable.\cite{pines}
The 
strength of the microscopic analysis lies in the fact that it yields more 
information than can be garnered at the phenomenological level. Coupling 
of a plasma mode to single-particle excitations give rise to Landau 
damping and yields a correction to classical dispersion that can be 
systematically calculated using many-body theory.

With a sharp edge, there are so few degrees of 
freedom that one cannot distinguish between collective modes and 
single-particle excitations. 
This leads to bosonization of one-dimensional fluids\cite{wen1,wen2,stone}
for which the single-particle properties can be uniquely derived
once we know the collective excitation spectrum. 

The situation changes when one realizes that the physical edge is often 
not a sharp boundary separating an incompressible bulk from its 
surroundings. Chklovskii {\it et al.}\cite{csg} have demonstrated that with 
the 
gate-confined 2DEG, the edge is a rather wide, smoothly varying region 
of density. While the idea of compressible edge had been actively adopted 
both experimentally and theoretically\cite{comp}
in the past few years, no attempt, to our knowledge, 
has been made to bridge it with the bosonization theory of the edge. 
A number of works have sprung up recently to address the effect of 
long-range Coulomb interaction on the properties of bosonized edge,
\cite{coulomb} but 
the geometry is still that of a sharp edge as in Wen's original paper. 
One interesting attempt to understand the dynamics of the compressible 
region was made by Aleiner and Glazman\cite{aleiner}, henceforth referred 
to as AG, who used classical hydrodynamics of 
charged fluid to derive the (necessarily collective) spectrum of the 
compressible edge analytically. They showed there are many density modes 
within the compressible edge, one of which is charged and higher in 
energy than the other, so-called acoustic modes, which are neutral.  
Their model, however, does not account for the boundary effect taking 
place at the interface of compressible and incompressible regions. We have
tried in this paper to understand the gapless spectrum of a wide edge 
more rigorously from the quantum-mechanical standpoint.

In Sec.\ II we will show how the phenomenological quantization 
approach of Wen 
can be generalized to the dynamics of compressible edge including its 
boundary 
with the bulk, and recover AG's result along the way. 
We also discuss the issue of bosonization in this kind of edge. In Sec.\ 
III, we start 
from the second-quantized Hamiltonian to address the same problem. It 
turns out that the density modes are strongly coupled because of the 
diffuse distribution of the electron occupation.
In Sec.\ IV, we discuss various consequences of such coupling, including 
its possible effect on the tunneling density of states.

\section{Phenomenological approach to edge dynamics}
The geometry 
of the sample is that of a semi-infinite Hall bar where the boundary 
separating the compressible and incompressible liquid is defined as $y=0$ 
and the compressible strip extends in the positive $y$ 
direction (fig.\ 1(a)).

In the presence of a relatively weak local electric field fluctuation and 
a strong external magnetic field, one can write
	$v_x=\delta\!E_y/B$ and 
	$v_y=-\delta\!E_x/B$
where the field is produced either by external potential or as a 
result of electron-electron interaction. In turn, the local field can be 
expressed as the gradient of the change in potential energy. To 
completely specify the motion, we need the following equations for the 
local density $n(x,y)$.

\ba {{\partial n}\over{\partial t}}+{{\partial n}\over{\partial x}}v_x+
       {{\partial n}\over{\partial y}}v_y & = & 0,\label{eq:conteq}  \\
      n(x,y\!+\!h(x,y,t),t) & = & n_0(y).\label{eq:denseq}
\ea
The first equation is the continuity equation  for the divergence-free 
velocity field,
$\partial v_x/\partial x\!+\!\partial v_y/\partial y\!=\!0$. The second 
one parameterizes the dynamics by identifying points on the 
density contour $n_0(y)$ with $(x,y\!+\!h(x,y))$. The situation is depicted 
in fig.\ 1(b). The boundary between the compressible and incompressible 
region is defined by 
$n(x,h(x,0,t),t)=n_0(0)=\nu/2\pi l_0^2$ where $l_0$ is the magnetic length.
We will set both this and Planck's constant, $\hbar$, equal to 1 in 
the rest of the section. 
If one imagined a collection of very long strings 
stacked along $y$-direction, each carrying a label $n_0(y)$, the density 
fluctuation can be understood in terms of the strings wiggling about  
equilibrium, but never so violently enough that they cross other strings.

One can deduce a set of identities by differentiating Eq.\ (\ref{eq:denseq}) 
with respect to $x,y$, and $t$, which can be put back into 
Eq.(\ref{eq:conteq}) to give

\be
      {{\partial n_0}\over{\partial y}}
\left({{\partial h}\over{\partial t}}-v_y+
       v_x {{\partial h}\over {\partial x}}\right)
\left(1\!+\!{{\partial h}\over{\partial y}}\right)^{-1}=0.
\label{eq:hydroeq1}\ee 
We have ${\partial n_0}/{\partial y}\!<\!0$ inside the compressible 
region. For small fluctuations, ${\partial h}/{\partial y}$ should be 
small so that it is only the middle term which vanishes identically.
By defining 
$h(x,y,t)\equiv \partial\eta(x,y,t)/\partial{x}$, one arrives at the 
equation of motion of the compressible edge

\be
     {{\partial^2\eta}\over{\partial{x}\partial{t}}} = 
        v_y-v_x{{\partial^2\eta}\over{\partial x^2}} 
      =  {1\over B}{d\over{dx}}
   V(x,y+{{\partial\eta(x,y,t)}\over{\partial x}},t).
\label{eq:hydroeq2}\ee
where $V(x,y+h)$ is the local potential(fluctuation) at a given time $t$. 
One can also check that varying the 
following action reproduces the above equation of motion.
 
\be
   S = \int {1\over 2}{{\partial n_0}\over{\partial y}}\!
              {{\partial\eta}\over{\partial x}}\!
                       {{\partial\eta}\over{\partial t}} dx dy dt
    -  \int {{\partial n_0}\over{\partial y}}\!
  \int_0^{{\partial\eta}/{\partial x}}
      eV(x,y+y') dy' dx dy dt
\label{eq:action}\ee
with $e$ being the electric charge.
Quantization comes from the commutator, 
\be
  [\eta(x,y),{{\partial n_0}\over{\partial y'}}h(x',y')]=
          i\delta(x-x')\delta(y-y').
\label{eq:commute}\ee
Note that results derived in eqs.\ (\ref{eq:hydroeq2})-(\ref{eq:commute}) 
reduce to known 
expressions for sharp edge by taking the limit 
 $\partial n_0(y)/\partial y=-(\nu/2\pi)\delta(y)$.

When one is solely interested in the dynamics {\it inside} the compressible 
edge, as AG were, one can regard $V(x,y)$ as Coulomb
potential due to the fluctuating density. In linear approximation,
$n(x,y)\!-\!n_0(y)\!=\!\delta\!n(x,y)\!\approx\! 
-(\partial n_0/\partial y)h(x,y)$,
and the Hamiltonian is given, up to quadratic order in density, by

\be
   H= {e^2\over{2\kappa}}\int
         {{\partial n_0}\over{\partial y}}\!
         {{\partial n_0}\over{\partial y'}}\!
         {{h(x,y)h(x',y')}\over\sqrt{(x-x')^2+(y-y')^2}} dxdydx'dy';
        \hspace{0.5cm}\kappa=4\pi\epsilon_{0}\epsilon
\label{eq:hydroH}\ee
One can expand the operator $h(x,y)$ in terms of its Fourier components in 
$x$,$0\!<\!x\!<\!L$. Each component $h_{q}(y)$ satisfies

\be
    i\dot{h}_q(y)={{2 e^2 q}\over\kappa}
       \int_0^\infty dy'{{\partial n_0}\over{\partial y'}}
    K_0(|qy-qy'|)h_q(y'). 
\label{eq:hqdot}\ee
where $K_0$ is the modified Bessel function of zeroth order. For an 
eigenmode $h_{q\alpha}(y)$ whose time dependence is  
$e^{-i\omega_{\alpha} t}$, it becomes a homogeneous integral equation.
We can exploit the monotone nature of the function $n_0(y)$ to use
$n_0$ as a coordinate, and remove the partial derivative inside the 
integral.
Hence we arrive at a Fredholm integral equation with a real, symmetric 
kernel, for which the eigenfunctions form a complete, orthogonal set and 
all eigenvalues are real.\cite{arfken} The kernel $K_0$ is expressed in 
terms of eigenvalues and eigenfunctions by
\be
    K_0(|qy(n_0)-qy(n_0')|)=
{\kappa\over{2e^2 |q|}}\sum_{\alpha=0}^{\infty}
\omega_{\alpha}(q)h_{q\alpha}(n_0)h_{q\alpha}(n_0').
\label{eq:kernel}\ee

The eqs.\ (\ref{eq:hqdot})-(\ref{eq:kernel}) summarize the results initially 
obtained by AG. When we have the
boundary  separating the incompressible and compressible(IC) region, 
density fluctuation in the compressible region 
necessarily implies some of the electrons on the 
incompressible side have to move as well. When they move, they are doing 
work against the field that confines them, and it costs additional 
energy. The amount of energy necessary is given by

\be
	H_{IC}={{\nu v_F}\over{4\pi}}\int h(x,0)^2 dx.
\label{eq:Hic}\ee
where $y\!=\!0$ indicates the IC boundary. 
Indeed, this was the only term that appeared in the Hamiltonian for a 
sharp edge, without the electron-electron interaction. 

The total Hamiltonian, $H_{tot}=H+H_{IC}$ is written
in the basis of eigenfunctions as

\ba
    h(x,y) & = & L^{-1/2}\!\sum_{q\alpha} c_{q\alpha}h_{q\alpha}(y)e^{-iqx}, 
                    \nonumber \\
    H_{tot} & = & \sum_{q\alpha}
\left({\omega_{\alpha}(q)\over 2q}+{\nu 
v_F\over{4\pi}}h_{q\alpha}^2(0)\right) c_{q\alpha}^{\dag}c_{q\alpha} 
     + \sum_{q,\alpha\neq\beta}{{\nu v_F}\over{4\pi}}
         h_{q\alpha}(0)h_{q\beta}(0)c_{q\alpha}^{\dag}c_{q\beta}. 
\label{eq:Htotal}\ea
The boundary term contributes a diagonal term which modifies the slope of 
the dispersion but, more importantly, there are off-diagonal terms which mix 
states 
that were thought to be orthogonal. The perturbation is quadratic, hence 
by rotating to a proper basis, one can again recover independent 
modes. One might as well think of Eq.\ (\ref{eq:kernel}) as containing the 
complete dynamical 
information, provided we use the eigenmodes and eigenvectors of the full
Hamiltonian. 
Quantization condition, Eq.(\ref{eq:commute}), becomes 
$[c_{q\alpha}^{\dag},c_{q\alpha}]=q$ and the Hamiltonian becomes a sum of 
oscillator modes whose energy levels are set by 
$\omega_{\alpha}(q)$. Before mixing, $\omega_0(q)\propto q|\log q|$ is the 
most dominant mode while $\alpha\ge 1$ are nearly linear. The boundary 
will mix these and the linear modes become weakly logarithmic. 

From the set of harmonic oscillator-type operators at our disposal, 
we can follow the standard recipe of bosonization in 1$d$
\cite{wen1,wen2,stone} 
and construct the following set of fermion operators at the IC boundary.

\be
  \psi^{\dag}_{\alpha}(x)\propto \exp\{\sum_{q>0}
\sqrt{{2\pi}\over\nu}{1\over q}(c_{q\alpha}e^{-iqx}-
                                    c^{\dag}_{q\alpha}e^{iqx})\}
\label{eq:fermibose}\ee
Now $\psi^{\dag}_{\alpha}(x)$ involves a coherent density change over the 
whole 
compressible region including the IC boundary and differs from Wen's 
construction in which an electron is created at the perimeter of the 
incompressible fluid.

\section{Microscopic approach to edge dynamics}

Our starting point is the Coulomb Hamiltonian, 

\be
 H = {e^2\over{2\kappa}}\int {1\over |r-r'|}  
\psi^{\dag}(r)\psi^{\dag}(r')\psi(r')\psi(r) d^{2}rd^{2}r'.
\label{eq:operatorH}\ee 
We are 
interested in the dynamics within the lowest Landau level, therefore we 
can expand $\psi(r)$ as follows:\cite{chamon}

\be
\psi(r)=(\sqrt{\pi}L)^{-1/2}\sum_k e^{ikx}e^{-{1\over 2}(y-k)^2}c_k.
\label{eq:psi}\ee
When it is substituted into Eq. (\ref{eq:operatorH}), we obtain

\ba
 H & \approx & {e^2\over{\kappa L}}\!\!\sum_{k_1 k_2 q}\!\!e^{-{1\over 2}q^2}
 \!\!K_0(|qk_1\!\!-\!\!qk_2|)\cdag_{k_1\!+\!{q\over 2}}c_{k_1\!-\!{q\over2}}
          \cdag_{k_2\!-\!{q\over2}}c_{k_2\!+\!{q\over2}}
              \nonumber \\
     & = & {1\over2}\sum_{q\alpha} 
         {\omega_{\alpha}(q)}e^{-{1\over 2}q^2}\Pdag(q)\Palpha(q),
          \nonumber \\
\Pdag(q) & = & {1\over\sqrt{|q|L}}\sum_{k} \hqa(k)\cdag_{k+q/2}c_{k-q/2}
             =\Palpha(-q).
\label{eq:factorH}\ea
The difference between the normal-ordered form, Eq.\ (\ref{eq:operatorH}), 
and the above Hamiltonian is an infinitely large constant which can be 
cancelled by assuming a compensating positive charge background.
The Hamiltonian is approximate in the sense that we have replaced the 
convolution $\int f(y)e^{-(y-k)^{2}/2}dy$ with
$\sqrt{2\pi}f(k)$ or, in other words, identified $y$ with $l_0^2 k$.
We have used the expansion of $K_0$, 
Eq.\ (\ref{eq:kernel}), 
and it should be understood that the effect of the IC 
boundary can be included by resorting to appropriate generalizations of 
$\omegalpha(q)$ and $\hqa$. 

More can be understood about the form of the above Hamiltonian by 
recalling that for a homogeneous system, the Coulomb Hamiltonian takes 
the form
$1/2\sum_{\vec{q}}V(\vec{q})\rho(\vec{q})\rho(-\vec{q})$, provided one 
discards the exchange term. Because the $y$-direction is not 
translationally invariant, we have $\alpha$ taking over the role of a 
momentum, labeling the underlying classical transverse modes. 

The hydrodynamic result of the previous section is tantamount to having 
each density operator, $P_{\alpha}^{\dag}(q)$, oscillating at a definite 
frequency, $\omega_{\alpha}(q)$. We will examine the Heisenberg equation 
of motion of the density operators as follows:

\be -i\dot{\Pbeta}(p)=
{1\over 2}\sum_{q\alpha}\omegalpha(q)\{\Pdag(q)[\Palpha(q),\Pbeta(p)]+
[\Palpha(q),\Pbeta(p)]\Pdag(q)\}.
\label{eq:Pdot}\ee
We have absorbed the overlap integral $e^{-q^2/2}$ into $\omegalpha(q)$.
The r.h.s. of the above equation contains the hydrodynamic term,
$\omega_\beta (p)\Pbeta(p)$, but there are also  off-diagonal 
components present. Evaluation of those terms must rely on some kind of 
approximation scheme. We will solve Eq.\ (\ref{eq:Pdot}) by linearizing 
the r.h.s. so that it reduces to a set of coupled linear equations for 
$\Pbeta(p)$. 

The first method we try is a version of RPA, in which one only keeps those 
bilinear terms with the momentum transfer equal to $p$, i.e. terms 
proportional to $\cdag_{k+p}c_{k}$.\cite{pines,mahan}
To leading order in momenta, we get

\be
\Pdag(q)[\Palpha(q),\Pbeta(p)]\approx \newline
\delta_{\alpha\beta}\delta_{pq}\Pbeta(p)-{2\over{p^{1/2}qL^{3/2}}}
\sum_k \hqa(p\hqa^{'}\hpb+q\hqa\hpb^{'})n_k\cdag_{k+p/2}c_{k-p/2}.
\label{eq:RPA}\ee
The primes indicate derivatives with respect to $k$. 
In deriving the above result, we have used 

\be
[\Palpha(q),\Pbeta(p)]\approx
(1/\sqrt{|pq|}L)\sum_k \{p\hqa^{'}\hpb+q\hqa\hpb^{'}\}
\cdag_{k+(p-q)/2}c_{k-(p-q)/2}.
\label{eq:Pab}\ee 
The diagonal term in Eq.\ (\ref{eq:RPA}), when multiplied by 
$\omegalpha(q)/2$ and summed over  $q$ and $\alpha$, 
gives half of $\omega_{\beta}(p)\Pbeta(p)$. The other half comes from 
$[\Palpha(q),\Pbeta(p)]\Pdag(q)$. 
In evaluating the second term in Eq.\ (\ref{eq:RPA}), we assume that 
various eigenfunctions do not depend on $q$ as much as they do on 
$\alpha$ so that we can simply label them $h_{\alpha}(k)$. The 
integration over $q$ extends from 0 to 
$\infty$, but due to rapidly decaying $e^{-q^2/2}$, there is a  
cutoff at large momenta. Since our interest lies in estimation, we 
conveniently choose the region of integration for $q$ to be 
$[0,p]$. The Eq.\ (\ref{eq:Pdot}) becomes

\be
 \omega_{\beta}(p)\Pbeta(p)
-\sum_{\alpha}\omegalpha(p){1\over{ (pL)^{1/2} }}\sum_k
 {p\over{2\pi}}(h^2_{\alpha}h_{\beta})^{'}n_k\cdag_{k+p/2}c_{k-p/2} =
\newline
  \omega_{\beta}(p)\Pbeta(p) 
 -\sum_{\alpha,\gamma}\omegalpha(p)A_{\alpha\beta\gamma}
 P^{\dag}_{\gamma}(p)
\label{eq:correction}\ee
where $A_{\alpha\beta\gamma}$ is a constant of order $l_0^2/aL$, $a$ 
being the width of the compressible edge. The size of coupling of
$\Pbeta(p)$ to $P^{\dag}_{\gamma}(p)$ is the sum 
$\sum_{\alpha}\omegalpha(p)A_{\alpha\beta\gamma}\sim
(l_0^2/aL)\sum_{\alpha}\omegalpha(p)\sim 
(l_0/L)\langle\omegalpha(p)\rangle_{avg}$. In arriving at the last 
relation, we assumed the net number of modes involved are roughly $a/l_0$.
So the coupling between modes is generally weak.

The underlying assumption of the above analysis, or indeed in the original 
RPA theory, is that there exists a {\it stable} collective 
excitation. For a stable excitation,
$p$ is a good quantum number, and it 
may be safe to keep just those terms that preserve $p$. In the theory of 
Nozi\`{e}res and Pines\cite{pines}, it is {\it a posteriori} justified by 
calculating various decay rates of a plasmon which turns out to be small 
compared to the 
energy scale of a plasmon. In our problem, there is no obvious energy scale 
to set collective modes apart from single particle excitations and there 
may be a large coupling between the two. This may in turn show up as a 
coupling between modes of different momenta and if it is significant, 
will destroy the assumption of the RPA.

As another way to linearize Eq.\ (\ref{eq:Pdot}), we propose to
replace the commutators with 
their expectation values in the ground state. The resulting 
equation will range over differing momenta as well as $\alpha$.
Taking the expectation value of Eq.\ 
(\ref{eq:Pab}) 
requires the knowledge of $\langle \cdag_{k+p}c_k\rangle$.
For $p\!=\!0$, it is just the density, $n_k$. For $p\!\neq\!0$, we can 
argue the 
following. If the state $k$ was unoccupied in the ground state, $c_k$ acting 
on the ground state would destroy it. Likewise, if the state $k\!+\!p$ had 
been 
occupied, there would be no place for an added electron to go, hence the 
expectation value will be again zero. So a plausible answer is 
$\langle\cdag_{k+p}c_k\rangle\!=\!n_k(1\!\!-\!\!n_{k+p})\!\approx\! 
n_k(1\!\!-\!\!n_k)$ for small $p$. We get 
precisely this result for the form of the ground state 
$|g.s.\rangle=\prod_k (\sqrt{1\!\!-\!\!n_k}+
\sqrt{n_k}\cdag_k)|vac\rangle$. There are as yet no definitive answers in 
the literature on how the 
ground state of the edge should behave at $T=0$. Our guess is that if 
one is at a sufficiently high temperature that any many-body correlations 
which may exist at absolute zero
are destroyed, but still low enough that quantum hall 
effect survives, the ground state should be approximately described by the
single particle state wavefunction given above. 

The commutator part can now be worked out straightforwardly. For $p=q$, 
it is equal to $\delta_{\alpha\beta}$. When $p\neq q$,

\be
\langle[\Palpha(q),\Pbeta(p)]\rangle=
{{p+q}\over{4\pi\sqrt{pq}}}S_{\alpha\beta}+
{{q-p}\over{4\pi\sqrt{pq}}}A_{\alpha\beta},
\label{eq:expPab}\ee
where $S_{\alpha\beta}=\int_0^{\nu}h_{\alpha}h_{\beta}(1-2n_k)dn_k$, and 
$A_{\alpha\beta}=\int_0^{\nu}(h_{\beta}\partial h_{\alpha}/\partial{n_k}\!\!-
\!\!h_{\alpha}\partial h_{\beta}/\partial{n_k})(n_k\!\!-\!\!n_k^2)dn_k$. The 
two integrals 
contribute numbers of order unity as shown in the Appendix. 
We find it more convenient to carry out the analysis in terms of
$P^{\dag}_{\alpha}(p)/\sqrt{p}$ and 
will refer to it as $P^{\dag}_{\alpha}(p)$. We write 
$\omega_{\alpha}(p)=v_{\alpha}p$ where it is understood that the velocities 
have weak logarithmic 
dependence and $v_0/v_1\!\approx \! -\log(pa)\!\approx\!\log 
10^4$.\cite{aleiner}
With $\nu\!=\!1$ and assuming the existence of just two modes, $\alpha=0,1$, 
we have

\ba
    -i\dot{P_0^{\dag}}(p) & = & \omega_{0}(p)P_{0}^{\dag}(p)+
{4\sqrt{2}\over{\pi^2}}\omega_{1}(p)\sum_q P^{\dag}_{1}(q) 
               \nonumber \\
   -i\dot{P_{1}^{\dag}}(p) & = & \omega_{1}(p)P_{1}^{\dag}(p)+
{4\sqrt{2}\over{\pi^2}}\sum_q \omega_{0}(q)P^{\dag}_{0}(q).
\label{eq:nu1}\ea
What we have now is a coupled equation without the presence of 
small dimensionless 
parameters, $l_0/a$, or $a/L$. This is the crucial feature that 
arises as a 
result of the microscopic calculation. The modes which were 
thought to be independent in the hydrodynamic model are in fact strongly 
coupled. 
One can do a perturbative analysis of the above equation using the fact that 
$\omega_1(p)/\omega_0(p)=v_1/v_0$ is small,

\ba
\dot{P_{1}^{\dag}}(p) & \approx &
 i{4\sqrt{2}\over{\pi^2}}\sum_q \omega_{0}(q)P^{\dag}_{0}(q) \approx
 \sum_q \dot{P_{0}^{\dag}}(q) \nonumber \\
-i\dot{P_0^{\dag}}(p) & \approx &
\omega_{0}(p)P_{0}^{\dag}(p)+
 \left({{4\sqrt{2}}\over{\pi^2}}\right)^2\!\!\omega_{1}(p)\sum_q\sum_{q'}
P^{\dag}_{0}(q').
\label{eq:perturb}
\ea
Up to this order of $v_1/v_0$, the dynamical behavior of $P_1^{\dag}(p)$ is 
completely determined by that of $P_0^{\dag}(p)$ and results in 
renormalization of the equation of motion for $P_0^{\dag}(p)$. 
With $\nu=1/3$, there is also coupling within the same mode, 

\ba
    -i\dot{P_0^{\dag}}(p) & = & \omega_{0}(p)P_0^{\dag}(p)+
   \sum_q {{p+q}\over{3}}v_{0}P_0^{\dag}(q) +
   \sum_q {\sqrt{2}\over 3}v_{1}(p-q+4p/\pi^2)
      P_{1}^{\dag}(q) \nonumber \\
    -i\dot{P_1^{\dag}}(p) & = & \omega_{1}(p)P_1^{\dag}(p)+
  \sum_q {{p+q}\over 3}v_{1}P_1^{\dag}(q) +
  \sum_q {\sqrt{2}\over 3}v_{0}(q-p+4q/\pi^2)
      P_{0}^{\dag}(q). 
\label{eq:nu1/3}\ea

Why is there so much mixing? For a sharp edge, the occupation number makes 
a step function, and we get $\langle\cdag_{k+p}c_k\rangle=\delta_{p0}n_k$. 
Compressible edge has a diffuse distribution of electrons, giving a 
fairly large contribution to $\langle\cdag_{k+p}c_k\rangle$ for $p$ 
different than zero, 
and hence to the coupling of modes. In the light of the above analysis, our 
initial RPA  assumption to keep only the momentum-conserving terms seems 
unjustifiable. 

\section{Discussion}

The density excitation at the edge is often called the edge 
magnetoplasmon (EMP) and to distinguish the different modes, the 
$\alpha\ge 1$ are called the ``acoustic" modes. The Eqs.\ (\ref{eq:nu1}) 
and (\ref{eq:nu1/3}) then represent the coupling of the EMP modes with the
acoustic modes. Since
experiments are often done at the fundamental frequency corresponding to 
$p=2\pi/L$, one can keep just $P_0(2\pi/L)$ and $P_1(2\pi/L)$ in Eqs.\ 
(\ref{eq:nu1}) and (\ref{eq:nu1/3}) and disregard higher modes. Note that 
the Gaussian factor implicit in the frequency $\omega_\alpha$ tends to 
suppress the coupling to higher momentum modes.
We also introduce damping of the acoustic mode,
$\omega_1(p)=(v_{1}+i\gamma)p$ where $\gamma$ is comparable or larger 
than $v_1$ but still quite small compared to $v_0$. This is the problem 
of two coupled harmonic oscillators one of which is heavily damped. The 
resulting frequencies are easily found to be $v_0\!\rightarrow\! 
v_0\!+\!\lambda^2(v_{1}\!+\!i\gamma)$ and $v_1\!\rightarrow\!
(1\!-\!\lambda^2)(v_{1}\!+i\!\gamma)$, with $\lambda=4\sqrt{2}\nu/\pi^2$.
As a result of coupling, the initially dissipationless mode becomes 
weakly dissipative, and the acoustic mode propagates more slowly. 
The acoustic modes, $\alpha\ge 1$, were predicted to be essentially
unobservable due to dissipation\cite{aleiner} in currently available samples
with mobilities $\mu\!\approx\! 100$ $\mbox{m}^2/\mbox{Vsec}$. A number of
EMP experiments both in the
integer\cite{talyanski} and fractional\cite{wass,ashoori} have been unable
to detect the diversity of modes either. Our analysis shows the mode 
coupling works in favor of slowing down the speed of propagation of the 
acoustic mode, thus making its detection more difficult.

For higher momentum states, the effect of coupling should be more 
significant, as one can see from the second half of Eq.\ 
(\ref{eq:perturb}). Since $\omega_1$ is complex, the more terms added in 
the summation will increase the dissipation. We therefore expect the states 
with large $p$ to decay easily compared to the modes close to the 
fundamental frequency.

   In the experimental study of Luttinger liquid behavior of the edge, 
it is necessary that 
the edge be as sharp as possible in order to both reduce the number of 
active modes in the edge, and to eliminate coupling between them.
Recent experiment of Chang {\it et al.}\cite{luttinger} was designed to 
create a very sharp edge profile, and the experiment seems to vindicate 
many predictions of the chiral Luttinger liquid theory. They see 
exponents in the I-V measurement which are not totally integral, as expected 
from the theory, but rather show deviations of the order of 10\%. In the 
light of this result, it is instructive to examine the possible 
correction of the tunneling density of states away from the ideal 
sharp-edge, short-range interaction limit. The long-range character of the 
dispersion 
makes the evaluation of the spectral function difficult, but it has been 
successfully solved in the recent work of Z\"{u}liche and 
MacDonald\cite{zuliche}. They conclude the long-range interaction results 
in the logarithmic correction to the tunneling density of states. We focus 
instead on the consequence of the coupling of modes. 

We assume a much simplified model with a single-mode edge 
excitation, such as given by $P_0^{\dag}(q)$ whose excitation is undamped 
for $p$ ranging from 0 to $p_c$. The modes with $p\!>\!p_c$ are damped with a 
constant coefficient, $\omega\!=\!p+i\gamma p (v_0\!=\!1)$. The source of 
damping will be the coupling to the invisible $P_1^{\dag}$ mode as discussed 
above. The density of states is 
extracted from the imaginary part of the retarded electron Green's function, 
$G_{R}(xt)\!=\!-i\langle\psi(xt)\psi^{\dag}(0)\rangle$, and the total 
tunneling 
density of states at energy $\omega$ is proportional to the imaginary 
part of 
$G_{R}(x=0,\omega)$. A straightforward calculation\cite{mahan} shows that 

\be
G_R(0,\omega)\propto
\omega^{m-1}\int_0^{\infty} \exp\{it+m\int_0^{p_c/\omega}
    {{dq}\over q}e^{iqt-q0^{+}}(1-e^{-q\gamma t})\} 
    {{dt}\over (t+i\gamma t-i0^{+})^{m}}
\ee
where $m=1/\nu$.
For $p_c=0$ (constant rate of damping for all momenta) the tunneling 
density of states is proportional to $\omega^{m-1}$, as is the case for a
non-dissipative system given by $\gamma\!=\!0$. With $p_c\neq 0$, the limit
of integration  explicitly depends on $\omega$ and the naive scaling 
result will be modified. For $p_{c}/\omega\!\ll\!1$, one can check  
that the first correction term goes like $(p_{c}+\omega)^{m-2}$. 
One should note, however, that other than this 
kind of dissipative correction and the effect of long-range interaction 
on the spectral density, the compressible edge will not pose an essential 
modification of the original sharp-edge theory.

\acknowledgments
  We are grateful to Ping Ao, Boris Spivak, Kiril Tsemekhman, Vadim 
Tsemekhman,  Carlos Wexler and Larry Yaffe for many useful discussions and 
feedback 
throughout the development of our work. This work was supported in part 
by the National Science Foundation under grant DMR-9528345.

\appendix\section*{Evaluation of $S_{\alpha\beta}$ and $A_{\alpha\beta}$}

One needs a specific expression of $h_{q\alpha}$ in order to evaluate 
$S_{\alpha\beta}$ and $A_{\alpha\beta}$. We follow AG and 
use
$h_{q0}(n_k)\!\!=\!\!\sqrt{2\pi/\nu}, 
   \hqa(n_k)\!\!=\!\! 2\sqrt{\pi/\nu}\cos(\pi\alpha n_k/\nu)$ for 
$\alpha\ge 1$.
One can check this is an exact set of solutions once the charged 
mode,$\alpha=0$, is so large in energy that it decouples from all of the 
lower modes. There is also mixing due to the IC boundary as mentioned in 
Sec.\ II, and the true eigenstates are a complicated mixture of the 
above set. We have not considered the mixing of states, but do not expect 
that it will significantly change the result that follows.

Using the above eigenfunctions, we obtain

\be
   S_{\alpha\beta}=S_{\beta\alpha}=\left\{\begin{array}{ll}
       \alpha=\beta, & 2\pi(1-\nu) \\
    \alpha=0\neq\beta, & (4\sqrt{2}\nu/\pi\beta^2)(1-(-1)^{\beta}) \\
   \alpha\neq\beta\neq 0, & (8\nu/\pi)
  \{(\alpha^2+\beta^2)/(\alpha^2-\beta^2)^2\}(1-(-1)^{\alpha+\beta})
   \end{array}\right.
\ee
and
\be
  A_{\alpha\beta}=-A_{\alpha\beta}=\left\{\begin{array}{ll}
   \alpha=0\neq\beta, & 2\sqrt{2}\pi\{
     (2\nu/\beta^2\pi^2)(1-(-1)^{\beta})-(1-\nu)(-1)^{\beta}\} \\
\alpha\neq\beta\neq 0, & 4\pi(1-\nu)
\{(\alpha^2+\beta^2)/(\alpha^2-\beta^2)\}(-1)^{\alpha+\beta}+ \\
     & (4\nu/\pi)  
\{(\beta-\alpha)/(\beta+\alpha)^3+
(\beta+\alpha)/(\beta-\alpha)^3\}(1-(-1)^{\alpha+\beta})
  \end{array}\right.
\ee
Some specific results for $\nu=1$ and $\nu=1/3$ are

$\begin{array}{llll}
      S_{00}=0=S_{11}, & S_{01}=8\sqrt{2}/\pi, & 
               A_{01}=8\sqrt{2}/\pi & (\nu=1)    \\                    
      S_{00}=4\pi/3=S_{11}, & S_{01}=8\sqrt{2}/3\pi, &
     A_{01}=4\sqrt{2}\pi/3+8\sqrt{2}/3\pi & (\nu=1/3) .
\end{array}$

\begin{figure}
\caption{Density profile given by $n$ of the edge in equilibrium, (a), and 
out of 
equilibrium, (b). The shaded region indicates incompressible bulk of filling 
factor $\nu$ and $a$ is the width of the edge. The constant 
density $n_0(y)$ is given by a straight, horizontal line 
in equilibrium, but is displaced by $h(x,y)$  at each point 
$(x,y)$ in (b). } 
\end{figure}


\begin{references}

\bibitem{wen1}X. G. Wen, Phys. Rev. B {\bf 41}, 12838 (1990);
X. G. Wen,  Phys. Rev. B {\bf 43}, 11025 (1991)
\bibitem{wen2}X. G. Wen, Int. J. Mod. Phys. B {\bf 6}, 1711 (1992).
\bibitem{haldane}F. D. M. Haldane and Yong-Shi Wu, \prel {\bf 55}, 2887 
(1985).
\bibitem{ao} Ping Ao and D. J. Thouless, \prel {\bf 70}, 2158 (1993).
\bibitem{pines} D. Pines and P. Nozi\`{e}res, {\it The Theory of Quantum 
  Liquids} (Benjamin, Reading, Mass., 1966).
\bibitem{stone} M. Stone, Ann. Phys. {\bf 207}, 38 (1991).
\bibitem{csg}D. B. Chklovskii, B. I. Shklovskii, and L. I. Glazman, Phys.
Rev. B {\bf 46}, 4026 (1992).
\bibitem{comp} P. L. McEuen {\it et al.}, \prbee {\bf 45}, 11419 (1992);
C. W. J. Beenakker, \prel {\bf 64}, 216 (1990); A. M. Chang, 
Solid State Commun. {\bf 74}, 871 (1990);
L. Brey, J. J. Palacios, and C. Tejedor, \prbee {\bf 47},
13884 (1993);
S. W. Hwang, D. C. Tsui, and M. Shayegan, \prbee {\bf 48}, 
8161 (1993);
B. Y. Gelfand and B. I. Halperin, \prbee {\bf 49}, 
1862 (1994);
M. M. Fogler, E. I. Levin and B. I. Shklovskii, \prbee {\bf 49}, 13767 
(1994); M. R. Geller and G. Vignale, \prbee {\bf 50}, 11714 (1994).
\bibitem{coulomb} Y. Oreg and A. M. Finkel'stein, \prel {\bf 
74}, 3668 (1995); K. Moon and S. M. Girvin, preprint (1995).
\bibitem{aleiner} I. L. Aleiner and L. I. Glazman, Phys. Rev. Lett.
 {\bf 72}, 2935 (1994)
\bibitem{arfken} G. Arfken, {\it Mathematical methods for 
physicists} (Academic Press, New York, 1985), p.890-893.
\bibitem{chamon} C. de C. Chamon and X. G. Wen, \prbee {\bf 49}, 8227 (1994).
\bibitem{mahan} G. D. Mahan, {\it Many-particle Physics} (Plenum Press, 
New York, 1981). 
\bibitem{talyanski} V. I. Talyanskii {\it et al.}, \prbee {\bf 50}, 1582 
(1994); V. I. Talyanskii {\it et al.}, J. Phys.: Condens. Matter {\bf 7}, 
L435-L443 (1995). 
\bibitem{wass} M. Wassermeier {\it et al.}, Phys. Rev. B {\bf 41}, 10287
 (1990).
\bibitem{ashoori} R. C. Ashoori {\it et al.}, Phys. Rev. B {\bf 45}, 
                   3894 (1992).
\bibitem{luttinger} A. M. Chang, L. N. Pfeiffer, and K. W. West, preprint.
\bibitem{zuliche} U. Z\"{u}liche and A. H. MacDonald, cond-mat/9607075. 
\end{references}
\end{document}